\documentclass[epsfig]{aipproc}
\layoutstyle{8x11double}

\begin{document}

\def\gr{\hbox{ \raisebox{-1.0mm}{$\stackrel{>}{\sim}$} }}
\def\kr{\hbox{ \raisebox{-1.0mm}{$\stackrel{<}{\sim}$} }}

\title{Mid-infrared observations of the SGR 1900+14 error box}


\author{Sylvio Klose}{
  address={Th\"uringer Landessternwarte, 07778 Tautenburg, Germany},
  email={klose@tls-tautenburg.de},
  thanks={}}

\author{Bringfried Stecklum}{
  address={Th\"uringer Landessternwarte, 07778 Tautenburg, Germany},
  email={stecklum@tls-tautenburg.de},
  thanks={}}

\author{Dieter H. Hartmann}{
  address={Clemson University, Clemson, South Carolina, SC 29634-0978},
  email={hartmann@grb.phys.clemson.edu},
  thanks={}}

\author{Frederick J. Vrba}{
  address={US Naval Observatory, Flagstaff, AZ 86002-1149},
  email={fjv@nofs.navy.mil},
  thanks={}}

\author{Arne A. Henden}{
  address={USRA/US Naval Observatory, Flagstaff, AZ 86002-1149},
  email={aah@nofs.navy.mil},
  thanks={}}

\author{Aurore Bacmann}{
  address={Astrophysikalisches Institut und Universit\"ats-Sternwarte, 
           07745 Jena, Germany},
  email={bacmann@astro.uni-jena.de},
  thanks={}}

\copyrightyear  {2001}


\begin{abstract}
We report on mid-infrared observations of the compact stellar cluster
located in the proximity of SGR 1900+14, and the radio/X-ray position of 
this soft-gamma repeater. Observations were performed in May and June of
2001 when the bursting source was in an active state. At the known radio
and X-ray position of the SGR we did not detect transient mid-IR activity, 
although the observations were performed 
only hours before and after an outburst in the high-energy band.
\end{abstract}


\date{\today}

\maketitle

                   \section{Introduction}

Recent deep, high-resolution multi-wavelength timing observations of 
SGRs led to significant progress in our understanding of these 
enigmatic sources [1, 2]. Key goals in current SGR studies include the 
identification of their counterparts at long-wavelengths and a better
understanding of their past and future evolutionary states (e.g., [3]). 
Of particular interest is their possible relation to the class of 
anomalous X-ray pulsars [4].
 
From the point of view of ground-based astronomy, among the known four
(perhaps five) soft gamma-ray repeaters [1] SGR 1900+14 has the
advantage that it (or better, its error box) is observable from
the northern as well as the southern hemisphere. This increases
the opportunities to monitor this source in the optical/infrared
bands whenever it is in an active state. During its recent activity
cycle in spring/summer 2001 we observed the SGR 1900+14
error box with the ESO 3.6-m telescope using the newly
commissioned TIMMI~2 mid-infrared camera. The campaign covered the
position shortly after and before an outburst in the high-energy band. 

                   \section{Observations}

TIMMI~2 (Thermal Infrared Multi Mode Instrument) is a thermal infrared
camera designed for direct imaging at 5, 10, and 17 microns. This instrument 
is the successor of TIMMI~1 which was decommissioned in
1999. TIMMI~2 uses a 240$\,\times\,320$ pixel
AsSi BIB detector [5], the scale is 0.3 arcsec per pixel for observations
in the $N$ band. The field of view is $96''\,\times\,72''$. An
overview of TIMMI~2 is given in [6], and some early observational results
are presented in [7].

Our first observing run was performed on May 22, about 1 month after
the giant gamma-ray flare detected from SGR 1900+14 on April 18 [8],
and also about 1 month before the detection of the next high-energy
outburst from this source [9]. A second observing run, again using
TIMMI~2 at the ESO 3.6-m telescope, was carried out on June 28, now
only about 9 hours after and 6.5 hours before a gamma-ray outburst
[9, 10]. All observations were performed using the N11.9 filter.
N11.9 is a narrow-band filter centered at about 11.6 microns (FWHM
of $\sim$ 1.2 microns). We selected this filter, because it offers the
highest sensitivity [5].

               \section{Results and Discussion}

With a flux density limit of about 3 mJy our observations
represent the deepest mid-infrared observations of the SGR 1900+14
field performed to date. Furthermore, with a delay of only 9
hours after a high-energy outburst our observations probe the SGR
environment at a time when it could still be affected by the energy 
input from the burst. However, we do not detect any mid-IR flux that
would indicate an energizing interaction between the burst and the
circum-burster medium. 

Basically, there are two issues that can be addressed with our
observations. The first one is the relevance of the compact stellar
cluster of high-mass stars seen in the proximity (Fig.~1) to the
X-ray/radio position of SGR 1900+14 [13]. The second issue concerns
the implications of the non-detection of the SGR in the mid-IR for
models of the burst source and its immediate environment.

\begin{figure}[t!]
\resizebox{18pc}{!}{\includegraphics{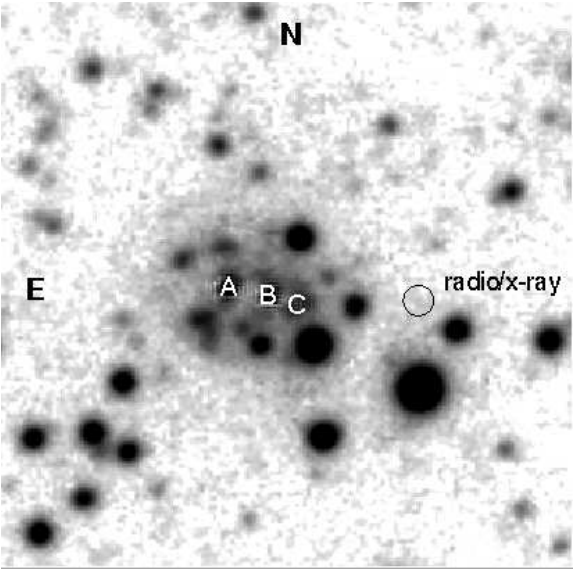}}
\caption{Deep $I$-band image of the
compact stellar cluster in the proximity of SGR 1900+14
with the position of its two most prominent members, two nearly
identical M5 supergiants,
indicated as A,B. In order to identify the cluster the M5 supergiants
and star C have been psf subtracted. 
Also indicated is the putative radio/X-ray position of SGR 1900+14  
suggested in [11, 12]. Adapted from [13]. To provide a scale, star B is
separated by 11.8 arcsec from this radio position [14].}
\end{figure}

\subsection{The compact stellar cluster}

To our knowledge, the only mid-infrared observations of the SGR
1900+14 error box to date were performed by van Paradijs et al. in July
1995 [15]. These authors used the ESO 3.6-m telescope equipped with
the TIMMI~1 camera. At the time of their observations neither a radio
transient nor an underlying compact stellar cluster was known, so
their main focus was the bright M5 supergiants
discovered in the arcmin-sized SGR error box ([16, 17]; Fig.~1).

\begin{figure}[t!]
\resizebox{18pc}{!}{\includegraphics{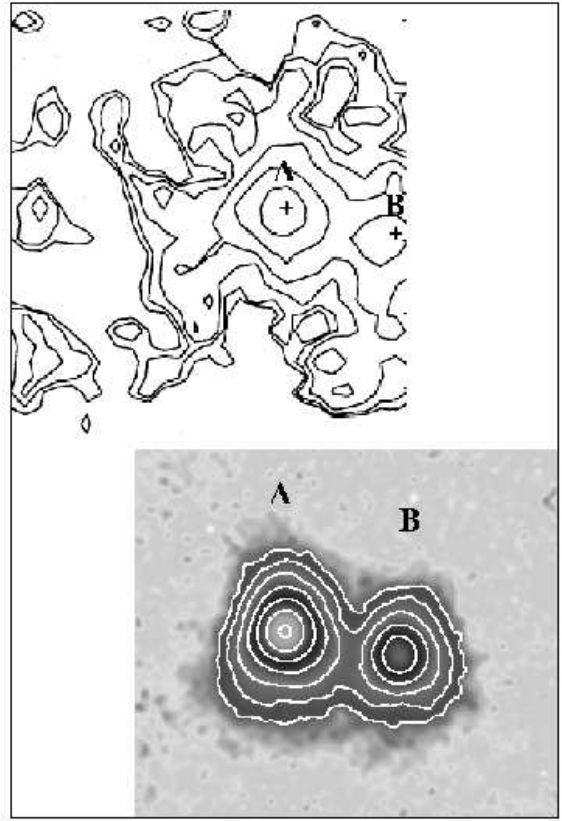}}
\caption{Part of van Paradijs' published countour plot of their
mid-infared observations with TIMMI~1 (top; [15]) compared to our
TIMMI~2 image (May 2001; bottom). The stars A and B refer to Fig~1. 
Contour levels correspond to 28, 56, 112, 224, 448 and 896 mJy per
square arcsec (top panel) and also include 14 mJy in the bottom panel. 
The point-spread function of the 3.6-m telescope is known to deviate 
slightly from circular.}
\end{figure}

Based on the earlier observations, our primary attention focused on 
potential evidence for long-term variability in the interstellar medium
surrounding the bursting source, including any gas associated with the M5
supergiants (Fig.~2). Van Paradijs et al. measured a flux density of
stars A and B  in the $N$ band of about 1.64 and 0.86 Jy, respectively
[15]. On our May 22 image we measure a flux density in the N11.9
filter of 2.11 and 0.90 Jy, respectively. Since the spectrum of these
stars is rapidly rising in the mid-infrared (see figure 5 in [17]),
these measurements are not in conflict with each other. Furthermore, they 
are in agreement with recent observations  of this region by the MSX
satellite during its Galactic Plane survey (Fig.~3; see the MSX database at
[18]): For the D-band image taken at 12.13 microns we measure a total flux 
density of the unresolved A+B stars of 2.97 Jy.
 
A comparison between our two observing runs seems to reveal a
slight short-term variability of star B. We consider this not
surprising, however, since supergiants are known to exhibit some
optical variability [17]. Spots on the stellar surface, for example,
could be responsible for variations of a cool supergiant [19]. Van
Paradijs et al. found evidence for a possible component of extended
diffuse emission surrounding the M5 supergiants [15]. Such a
component is not apparent in our data, although our observations are
more sensitive. The non-detection of a diffuse emission component is
interesting because it constrains the amount of hot dust within
the stellar cluster. Moreover, a large local flux of UV photons could
produce emission from PAHs of which one feature falls into our chosen
filter band. We do not detect this feature. Although we have not yet
performed detailed simulations, we believe that this non-detection indicates 
that there are no strong UV sources within this cluster. This is in
agreement with the non-detection of radio continuum emission from this
cluster [20]. Obviously this is a stellar cluster where high-mass
star-formation stopped \gr10$^7$ years ago.

\begin{figure}[t!]
\resizebox{18pc}{!}{\includegraphics{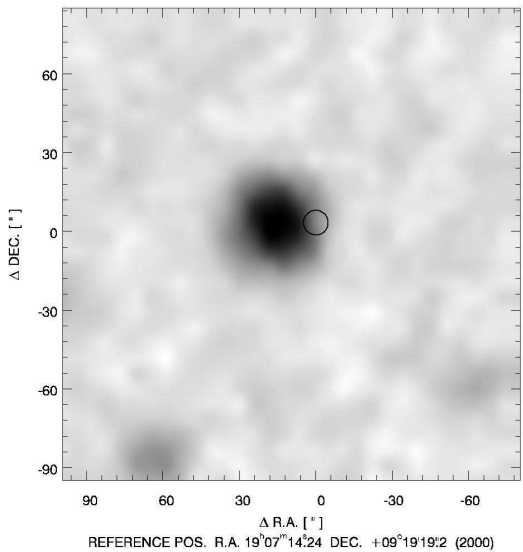}}
\caption{The SGR 1900+14 error box was also imaged by the Midcourse
Space Experiment (MSX satellite) in 1996/97 [18]. Shown is the
image obtained at 8.28 microns. Note that the  MSX satellite did
not resolve the stellar cluster. The image appears extended along the
scan direction of the satellite.}
\end{figure}

The recent discovery of a compact stellar cluster underlying the M
supergiants [13], and the discovery of a similar stellar cluster
close to SGR 1820--20 [21], raises the question of what role such 
clusters might play in the formation of SGRs. In the case of SGR
1900+14 there is an extensive debate in the literature of whether the
cluster and the SGR are indeed physically related or whether they are
located at very different distances in the Galactic Plane and only 
appear to be connected by random projection. Recently, the
debate  has focused on the different interstellar extinction measured
towards the M supergiants on the one hand [17], i.e. the most luminous
members of the stellar cluster, and toward the quiescent X-ray
counterpart of the SGR on the other hand [20]. 
Since the former is several magnitudes higher
than the latter, this seems to argue against a physical
relationship between the cluster and the SGR.

In principle, mid-infrared observations could be used to constrain the
distance of the stellar cluster, because they are less affected by
interstellar extinction. Assuming that star C in Fig.~1 is either 
a member of this stellar cluster or seen in front of this cluster,
we can place a
lower limit on the distance of the cluster by the fact that we do not
detect this star on our frames. The spectral type of this star was
determined by Vrba et al. to be M2 III [17]. Its predicted $N$-band
luminosity [22, 23] together with our non-detection of it 
places this star at a distance of $d \gr 8$ kpc, in agreement with
constraints deduced from earlier optical obervations [17]. A similar 
derivation of an upper limit on the cluster distance 
based on our observations is more uncertain.
Although the spectral type of the supergiants seems to be relatively
secure (M5; [17, 24]), there is  still the question of their
absolute luminosity in the $N$ band. Although theoretical $V-N$ colors
are available in the literature, the observational basis of
mid-infrared observations of supergiants is still very small. This
makes it difficult to estimate the uncertainty that can be attributed
to the choice of a certain $N$-band luminosity for these stars.

Although we cannot provide a strong upper limit on the distance of the
stellar cluster based on our mid-infrared observations, we draw  
attention to the following issue concerning the extinction measured
toward the M supergiants [17]. In light of the recent discovery of an
underlying compact stellar cluster [13] the possibility remains that
the measured extinction is the sum of ordinary
interstellar extinction and intrinsic extinction within the compact
stellar cluster. The observed scatter in the $I-J$ colors of the cluster
stars [13] significantly exceeds their uncertainties, indicating that 
several stars in this cluster suffer significant intracluster extinction. 
Not only that, but the spread of the $(I-J)$ color of the cluster stars
takes up about one half of the $\Delta (I-J)$ between the supergiants
A and B and the least reddened star at about $I-J \approx 6.7$ (table
1 in [13]). Naturally, there is no \it ad hoc \rm reason to assume
that the M supergiants, the brightest beacons of this cluster, are
by chance located in front of the stellar cluster. We discuss this point
further elsewhere [25].

\subsection{The SGR}

Naturally, our main hope was a possible detection of the SGR in the
mid-IR with observations placed so closely  
before \it and \rm after a high-energy outburst.
However, at the position of the radio transient associated
with the 1998 August 27 burst [11] as well as the quiescent X-ray
source discovered in the SGR  error box [12] we do not detect the
burster. The fact that there is no detection bears on the unknown
nature of the SGR environment. Because of the possible presence of
fossil accretion disks around SGRs, and their potential AXP relatives,
we modeled the spectral
energy distribution (SED) of such a disk following Perna et al. [26].  
At first glance, mid-infrared observations are very promising for 
detecting such disks since their SEDs can peak in the infrared. However,
even if we include 10 to 20 magnitudes of optical interstellar extinction
($A_V$) toward SGR 1900+14, compared to deep near-infrared
observations [20] our flux density limit does not provide a strong
constraint on any persistent accretion disk. We can report, however,
that if any such a disk exists around SGR 1900+14 then hours
before and after a high-energy burst its flux density in the $N$ band
does not exceed 3 mJy.

                   \section{Concluding Remarks}

The issue of whether or not SGR 1900+14 and the compact stellar cluster 
its proximity are physically related is crucially linked to the very
uncertain distances. Are these objects located at the same distance
or not? In this particular case, this question could be answered if a
better understanding of the measured extinction toward the M 
supergiants and toward the SGR is achieved. The former might profit
from deeper mid-infrared observations, the latter might gain from the
recent discovery [27] of a persistent dust-scattered X-ray halo
around the quiescent X-ray counterpart of the SGR, which could lead to
a direct measurement of the extinction by the scattering dust along
the line of sight [28].

The non-detection of any signal from the SGR might not be surprising
if there is no accretion disk at all around the burster, as indicated 
by the observations performed to date [20]. But should one expect to 
detect any non-gamma-ray signal from the burster within hours of a
high-energy outburst? This question remains to be addressed by further
theoretical studies, but we note that there are now two cases where mid-IR
observations were performed only a few hours after a SGR outburst
(the other case is SGR 1820--20 [21]) and no signal from the burster
or the ambient interstellar medium was detected down to a flux density
limit of a few mJy. Future observations need to further push the 
sensitivity limit, and also sample more closely in time. Truly 
simultaneous coverage would require robotic observations, similar 
to those used in the search for prompt optical emission [29, 30].

\begin{theacknowledgments}
The authors thank Jochen Greiner and U.R.M.E. Geppert (both AIP Potsdam, 
Germany) for valuable comments on the manuscript.
\end{theacknowledgments}



\begin{thebibliography}{12}

\bibitem{[1]}  Hurley, K., in \it Gamma-Ray Bursts\rm, edited by
               R. M. Kippen, R. S. Mallozzi, and G. J. Fishman, 
               AIP Conference Proceedings 526, American Institute of Physics,
               New York, 2000, p. 763.
\bibitem{[2]}  Ibrahim, A. I. et al., \it 
               Astrophys. J., \bf 558\rm, 237 (2001).
\bibitem{[3]}  Kaplan, D. L. et al., \it 
               Astrophys. J., \bf 556\rm, 399 (2001).  
\bibitem{[4]}  Thompson, C. et al., \it 
               Astrophys. J., \bf 543\rm, 340 (2000).  
\bibitem{[5]}  TIMMI~2 users manual, http://www.ls.eso.org/lasilla/
               Telescopes/360cat/timmi/html/manual.html.
\bibitem{[6]}  Relke, H. et al., \it SPIE \rm Vol. \bf 4009\rm, 440.
\bibitem{[7]}  Stecklum, B., in \it The Origins of 
               stars and planets: The VLT view\rm, edited by
               J. Alves et al., ESO, Garching 2001, in press. 
\bibitem{[8]}  Guidorzi, C. et al., GCN \#1041 (2001).
\bibitem{[9]}  Ricker, G. et al., GCN \#1073 (2001). 
\bibitem{[10]} Ricker, G. et al., GCN \#1074 (2001). 
\bibitem{[11]} Frail, D. A., Kulkarni, S. R., and Bloom, J. S., 
               \it Nature \bf 398\rm, 127 (1999).
\bibitem{[12]} Fox, D. W. et al., astro-ph/0107520 (2001).
\bibitem{[13]} Vrba, F. J. et al., \it Astrophys. J., \bf 533\rm, L 17 (2000).
\bibitem{[14]} Vrba, F. J. et al., \it Gamma-Ray Bursts\rm, edited by
               R. M. Kippen, R. S. Mallozzi, and G. J. Fishman, 
               AIP Conference Proceedings 526, American Institute of Physics,
               New York, 2000, p. 809.
\bibitem{[15]} van Paradijs, J. et al., \it Astron. Astrophys.,
               \bf 314\rm, 146 (1996).
\bibitem{[16]} Hartmann, D. H. et al., in \it Workshop on High Velocity
               Neutron Stars\rm, edited by R. E. Rothschild, and R. E. 
               Lingenfelter, AIP Conference Proceedings 366, 
               American Institute of Physics, New York, 1996, p. 84.
\bibitem{[17]} Vrba, F. J. et al., \it Astrophys. J., \bf 468\rm, 225 (1996).
\bibitem{[18]} see: http://www.ipac.caltech.edu/ipac/msx/
\bibitem{[19]} Tuthill, P. G., Haniff, C. A., Baldwin, J. E., \it 
               MNRAS \bf 285\rm, 529 (1997).  
\bibitem{[20]} Kaplan, D. L. et al., astro-ph/0107519 (2001).
\bibitem{[21]} Fuchs, Y. et al., \it Astron. Astrophys.,
               \bf 350\rm, 891 (1999).
\bibitem{[22]} Wainscoat, R. J. et al., \it Astrophys. J. Suppl. Ser. 
               \bf 83\rm, 111 (1992).  
\bibitem{[23]} Ducati, J. R. et al., \it 
               Astrophys. J., \bf 558\rm, 309 (2001).  
\bibitem{[24]} Guenther, E., Klose, S., and Vrba, F. J., in
               \it Gamma-Ray Bursts\rm, edited by
               R. M. Kippen, R. S. Mallozzi, and G. J. Fishman, 
               AIP Conference Proceedings 526, American Institute of Physics,
               New York, 2000, p. 825.
\bibitem{[25]} Klose, S. et al., in preparation (2002).
\bibitem{[26]} Perna, R., Hernquist, L., and Narayan, R.,
               \it Astrophys. J., \bf 541\rm, 344 (2000).    
\bibitem{[27]} Kouveliotou, C. et al., \it Astrophys. J., 
               \bf 558\rm, L 47 (2001).
\bibitem{[28]} Predehl, P., and Klose, S., \it Astron. Astrophys.,
               \bf 306\rm, 283 (1996).
\bibitem{[29]} Akerlof, C. et al., \it Astrophys. J.,  
               \bf 542\rm, 251 (2000).
\bibitem{[30]} Park, H. S. et al., \it Astron. Astrophys. Suppl. Ser., 
               \bf 138\rm, 577 (1999).

\end{thebibliography}
\end{document}